# Achieving ultra-high anisotropy in thermal conductivity of plastic crystal through megapascal pressure of hot pressing

Zhipeng Wu[1, #], Mingzhi Fan[3, #], Yangjun Qin[1, #], Guangzu Zhang[3, †], Nuo Yang[1, 2, †]

1) School of Energy and Power Engineering, Huazhong University of Science and Technology, Wuhan 430074, China

2) Department of Physics, National University of Defense Technology, Changsha 410073, China

3) School of Intergrated Circuits, Huazhong University of Science and Technology, Wuhan 430074, China

\# Z. W., M. F. and Y. Q. contributed equally to this work.

†Corresponding E-mail: G.Z. (zhanggz@hust.edu.cn) and N.Y. (nuo@nudt.edu.cn, nuo@hust.edu.cn)

# ABSTRACT

Plastic crystals, owing to their exceptional properties, are gradually finding applications in solid-state refrigeration and ferroelectric fields. However, their inherently low thermal conductivity restricts their utilization in electronic devices. This study demonstrates that applying megapascal pressure of hot pressing can enhance the thermal conductivity of plastic crystal films. Most importantly, it induces significant anisotropy in thermal conductivity. Such anisotropy in thermal conductivity is beneficial for specialized thermal management applications, such as directing heat flow paths in electronic devices. In this study, $[(CH_3)_4N][FeCl_4]$ PCs films were prepared by hot pressing. At a pressure of 16 MPa, the ratio of in-plane to cross-plane thermal conductivity in the film reaches a remarkable 5.5. This is attributed to the preferential orientation along the (002) crystal plane induced by uniaxial pressure, leading to the formation of a layered structure and the creation of a flat and dense film. Furthermore, according to molecular dynamics simulations, the thermal conductivity along the [100] and [010] directions (parallel to the (002) crystal plane) is higher than in other directions. Therefore, significant modulation of anisotropy in thermal conductivity is achieved in $[(CH_3)_4N][FeCl_4]$ films by applying uniaxial hot pressing pressure. This phenomenon has the potential to greatly broaden the application of plastic crystals in the field of flexible electronic devices.

# 1. Introduction

With the progress in solid-state refrigeration, plastic crystals (PCs) have emerged as highly promising materials due to the substantial phase transition entropy generated by the barocaloric effect[1-4]. PCs are a unique class of substances that exhibit characteristics between solid and liquid states, composed of rotatable molecules or ions[5, 6]. PCs exhibit outstanding mechanical deformability[7, 8] and hold great promise as solid-state refrigeration materials[1, 9, 10]. In the PCs systems, there exists a novel class of plastic/ferroelectric crystals that undergo a transition between plastic crystal phase and ferroelectric phase as the decreasing temperature[11]. $[(CH_3)_4N][FeCl_4]$ is a typical plastic/ferroelectric molecular crystal[11-13], showcasing multi-axis ferroelectricity and excellent plastic deformation capabilities. It can manifest ferroelectric polarization switching and significant piezoelectric response in the form of polycrystalline films. Consequently, PCs demonstrate significant research value in both the field of solid-state refrigeration and ferroelectricity.

However, PCs exhibit ultra-low intrinsic thermal conductivity at room temperature[14-17], which is in the order of 0.1 W/m-K, restricting its performance. The structure of PCs has a high disorder and the intermolecular forces are weaker than covalent bonding, leading to significant phonon scatterings[14, 18]. Additionally, PCs typically possess substantial compressibility or a small volume modulus[1], resulting in lower phonon group velocities and weaker intermolecular interactions[19, 20], thereby demonstrating low thermal conductivity. The lower thermal conductivity poses a significant limitation to its performance and restricts its application, especially in electronics[14]. Research indicates that 55% of electronic device failures result from the accumulation of heat leading to excessively high temperatures[21]. For electronic devices employing PCs, the low thermal conductivity may give rise to thermal hysteresis and could potentially affect their mechanical performance and charge-discharge rates[1, 3, 22]. The accumulation of heat may even induce phase transitions in the sample[11]. Therefore, modulating the thermal conductivity of PCs is crucial.

To modulate the thermal conductivity of PCs, methods typically involve altering their microstructure, which can enhance material crystallinity and molecular order, and even induce anisotropy in thermal conductivity. For common organic compounds, the methods such as mechanical stretching[23], electrospinning[24], nanoscale-templating[25], and hot pressing[26] can enhance the thermal conductivity and even induce anisotropy, enabling them to meet diverse heat dissipation requirements[27, 28]. Some studies have found through molecular dynamics (MD) simulations that altering the microstructure can enhance the thermal conductivity of organic polymers[29-31]. By mechanical stretching[32], it achieved a thermal conductivity as 62 W/m-K of polyethylene films in the in-plane direction. For organic ferroelectric materials, electric field polarization can increase molecular chain order, thereby enhancing thermal conductivity along the direction of polarization and generating thermal conductivity anisotropy[33]. Other methods, such as applying shear strain[34] and electrospinning[35], can also achieve high in-plane thermal conductivity and generate anisotropy. However, these methods typically present certain challenges in experimental implementation.

Due to the excellent mechanical deformability of PCs, the hot pressing method emerges as a straightforward and effective approach for fabricating films and modulating thermal conductivity[11, 12]. Hot pressing is a common method to prepare organic films, which can change the microstructure of the material[26]. In polymer composite systems, hot pressing can induce filler orientation[36], enhance in-plane thermal conductivity, and generating anisotropy[37, 38]. For PCs, a pressure field can also modulate their thermal conductivity. Wang et al. found that pressure suppresses the rotation of plastic crystal molecules, reduces resonance, and enhances in-plane thermal conductivity twice under a pressure of 3,500 MPa[39]. Li et al. observed a 10% enhancement in the thermal conductivity of neopentyl glycol under 200 MPa[40]. Xiong et al. observed that, for α-quartz, the thermal conductivity along the hexagonal direction only doubles under a pressure of 4 GPa. Furthermore, the ratio of thermal conductivity between the hexagonal direction and the basal direction increases from 1.5 to 2[41]. However, these studies present some questions worth exploring, such as whether an increase in in-plane thermal conductivity can be achieved without adding fillers, and whether achieving more pronounced anisotropy in thermal conductivity is feasible under lower pressures.

Here, it is investigated the impact of pressure during hot pressing on the thermal conductivity of PCs films. In this work, $[(CH_3)_4N][FeCl_4]$ films are prepared using the hot pressing method. It is found that under extremely low uniaxial pressure, the films exhibit exceptionally high anisotropy in thermal conductivity, with a significant increase in in-plane thermal conductivity. And the regulatory mechanisms are discussed by both microscopic structural morphology characterization and molecular dynamic simulations.

## 2. Experimental and simulation section

The preparation procedure for $[(CH_3)_4N][FeCl_4]$ is as follows: a specific quantity of iron(III) chloride hexahydrate is dissolved in hydrochloric acid and stirred to obtain solution A. Simultaneously, an equimolar amount of tetramethylammonium chloride is dissolved in deionized water and stirred to form solution B. Subsequently, solutions A and B are mixed and thoroughly stirred, resulting in a clear and transparent yellow solution. At 70°C, the solution undergoes rapid evaporation to yield powder samples for powder X-ray diffraction (XRD) testing. If the solution is left undisturbed for approximately 10 days, yellow crystals precipitate from the solution, constituting $[(CH3)_4N][FeCl_4]$. Next, PCs crystal grains are placed on a hot press, heating to 150°C, and maintaining at this temperature to prepare PCs films. Subsequently, pressure is applied and maintained for 20 minutes. Upon pressure release, a dense $[(CH_3)_4N][FeCl_4]$ film is formed. The films are then subjected to thermal conductivity testing, differential scanning calorimetry (DSC), XRD, scanning electron microscopy (SEM), and infrared temperature testing. Details of sample preparation are shown in supplementary material (SM) I.

The thermal conductivity of $[(CH_3)_4N][FeCl_4]$ films were measured using the 3-Omega method. Both the in-plane and the cross-plane value of thermal conductivities were obtained, where electrodes with two different widths were employed. Details of measurements are shown in SM II.

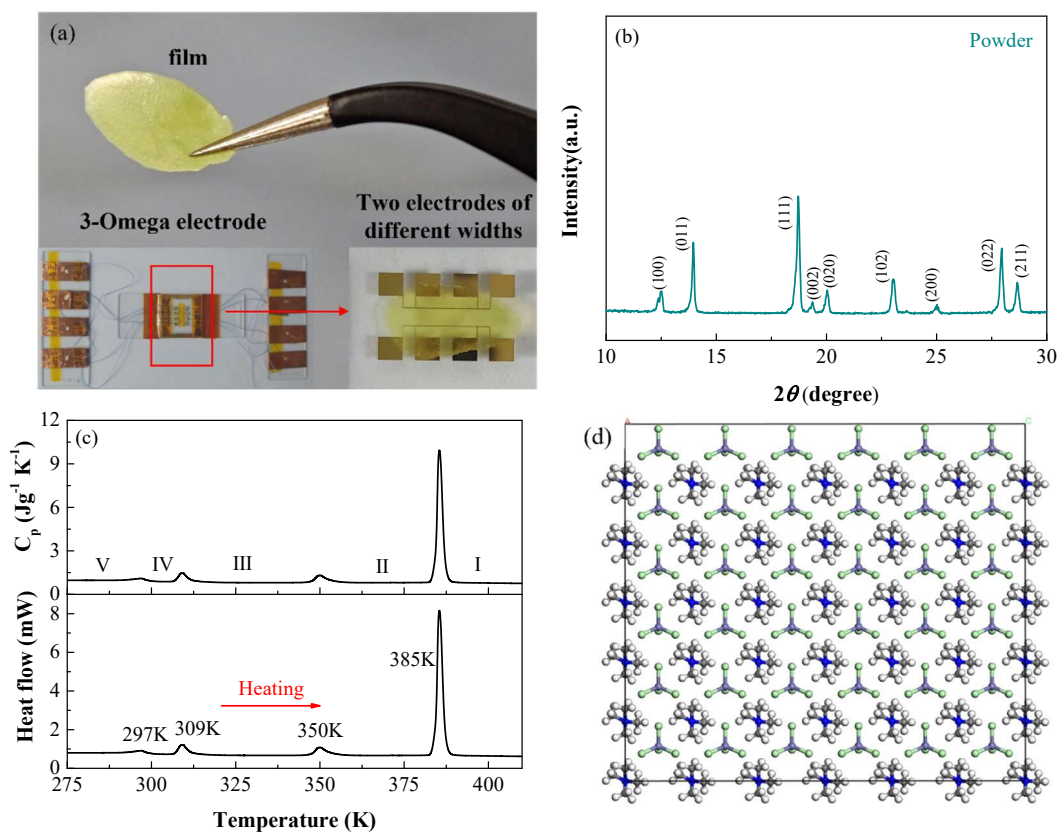

Figure 1: (a) Film sample and 3-Omega electrode images; (b) XRD pattern of plastic crystals powder; (c) Variation of heat capacity and heat flow with temperature in DSC testing; (d) Schematic representation of the Phase III structure of $[(CH_3)_4N][FeCl_4]$ molecules.

The sample images and 3-Omega electrode images of $[(CH_3)_4N][FeCl_4]$ are depicted in Figure 1(a). To confirm that the samples prepared in this study are $[(CH_3)_4N][FeCl_4]$, XRD and differential scanning calorimetry (DSC) tests were conducted on the samples. The XRD results in Figure 1(b) align with the XRD results of samples prepared by Harada et al[11]. Furthermore, DSC tests were performed on the thin film samples in this study. As illustrated in Figure 1(c), phase transitions occurred around 297 K, 309 K, 350 K, and 385 K, consistent with the reported transition points by Harada et al. These findings collectively indicate that the samples obtained in this experiment are $[(CH_3)_4N][FeCl_4]$.

The thermal conductivity of PCs was also calculated by MD simulations, which were conducted using LAMMPS software[42-44]. Employing the universal force field[45], a simulation cell of 2,780 atoms was performed. Then, thermal conductivities in different directions were obtained by equilibrium molecular dynamics. It was simulation different phases of $[(CH_3)_4N][FeCl_4]$, and the phase III structure was shown in Figure 1(d). Details of simulations are in SM III.

## 3. Results and discussion

The main finding of this study, shown in Figure 2(a), is that PCs films have a high anisotropy in thermal conductivity, after applying a lower pressure (>8 MPa) in hot pressing. For common inorganic materials like α-quartz and organic materials such as pentaerythritol[39, 41], a pressure at the GPa level is typically required to double the thermal conductivity and induce significant anisotropy in thermal conductivity. However, the pressure applied in this experiment is only at the MPa level. With increasing pressure, the in-plane thermal conductivity of the sample markedly increases, while the cross-plane thermal conductivity experiences a slight decline. The anisotropic ratio rises from 1.5 to 5.5 corresponding to the pressure increased from 4 to 16 MPa. At a pressure of 16 MPa, the in-plane thermal conductivity of the film reaches 0.59 ± 0.04 W/m-K, exhibiting a 146% increase compared to the in-plane thermal conductivity of 0.24 ± 0.04 W/m-K observed at 4 MPa pressure. Moreover, the highest ratio of in-plane to cross-plane thermal conductivity is observed as high as 5.5.

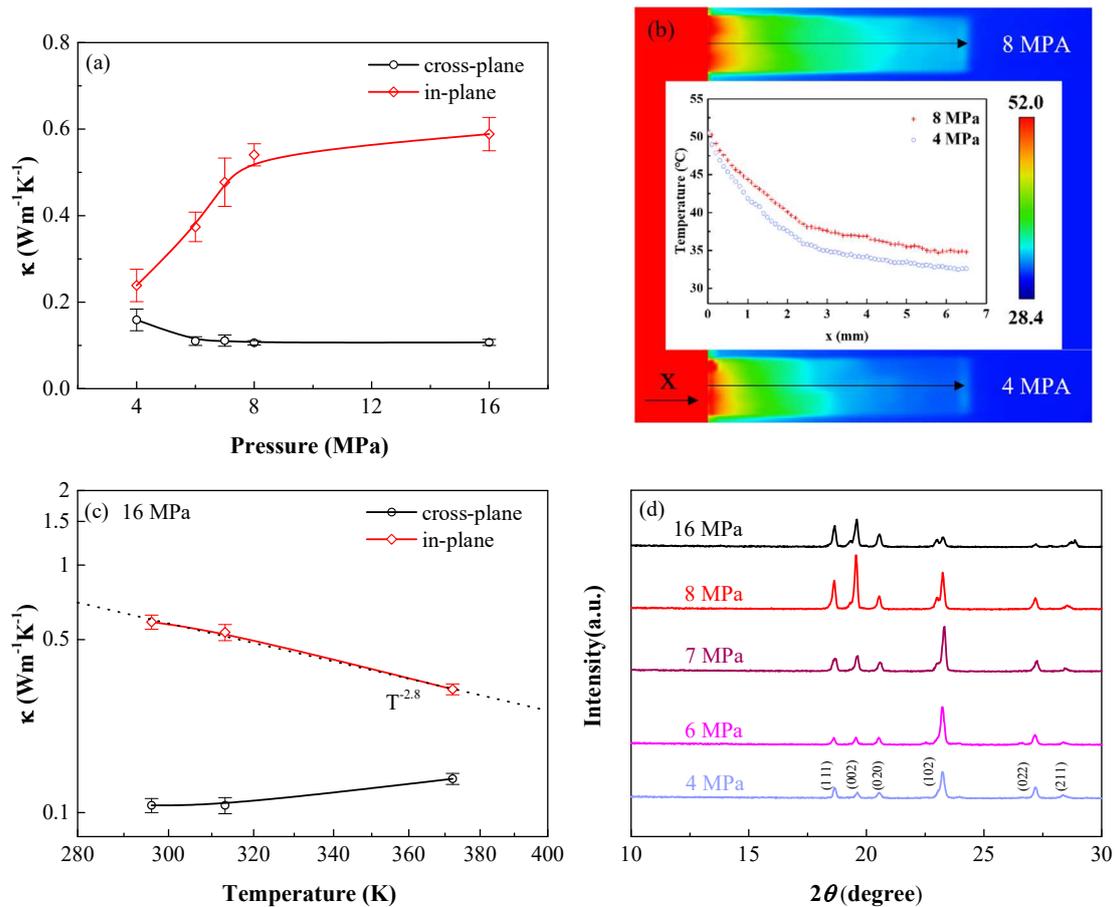

Figure 2: (a) Variation in thermal conductivity under different pressures (4 MPa, 6 MPa, 7 MPa, 8 MPa, or 16MPa); (b) Infrared thermographs of PCs films at different pressures under the same bottom temperature; (c) Variation in thermal conductivity under different temperatures; (d) XRD patterns of both plastic crystals powder and films prepared under varying pressures.

As shown Figure 2(b), temperature distributions in PCs films demonstrate the enhanced thermal conductivity by hot pressing pressure. The PCs films have dimensions of 6.5 mm in length, and 3.8 mm in width. The cross-sectional area of the films along the X-direction (length) remains constant. The films are placed on a temperature-controlled heating stage, with one end of the film in contact with the heated stage, while the bottom temperatures of both film ends are kept equal (52°C). The other end of the film is exposed to ambient temperature (28°C) without any additional internal heat sources. The convective heat transfer coefficient on the film surface is assumed to be constant, and the emissivity (ε) of the film is set to 0.70. The temperature variation along the x direction in the PCs films hot-pressed at 8 MPa is notably smaller. This effect arises because, for rectangular systems with identical cross-sections, higher thermal conductivity leads to reduced thermal resistance along x direction, resulting in a diminished temperature change along x direction. Therefore, this further substantiates that pressure indeed enhances the in-plane thermal conductivity of PCs films.

In Figure 2(c), it is shows that with increasing temperature, the anisotropic ratio of thermal conductivity in PCs films experiences a reduction. As shown in Figure 1(d), XRD characterization reveals that the peaks of powder are consistent with the Phase III structure of $[(CH_3)_4N][FeCl_4]$[11]. And due to the presence of thermal hysteresis, the sample is not in Phase IV at room temperature. Figure 2(c) shows that as the temperature increases, the in-plane thermal conductivity decreases, exhibiting a $T^{-2.8}$ relationship with temperature. Additionally, the anisotropy in thermal conductivity of the film decreases with increasing temperature. At elevated temperatures, the phonon vibrations within the lattice become more intense, leading to increased phonon-phonon scatterings, contributing to a noticeable decline in thermal conductivity. Moreover, due to the short-range disordered nature of $[(CH_3)_4N][FeCl_4]$, higher temperatures induce more pronounced molecular rotational motions, diminishing its initial orientational alignment[11, 46], and consequently reducing the anisotropy. Notably, at 372 K, the cross-plane thermal conductivity of the film increases from $0.11 \pm 0.01$ W/m-K to $0.14 \pm 0.01$ W/m-K, compared to 313 K. This is because the thermal conductivity of Phase II is higher than that of Phase III, and at 372 K, the orientation becomes more disordered, resulting in some orientations with higher thermal conductivity along the cross-plane direction (details will be discussed in the subsequent section on MD simulations).

It is worth noting that the property of ultra-high anisotropy in thermal conductivity can greatly expand applications of PCs[27, 28]. The ability to possess ultra-high thermal conductivity along specific planes or directions enhances the suitability of flexible films for specific electronic heat dissipation scenarios, ensuring reliable operating temperatures. Hence, it is essential to investigate the mechanism underlying the anisotropy in thermal conductivity of PCs films.

One of the potential reasons for the anisotropy is the preferential orientation of the crystal structure. The XRD pattern of the films are illustrated in Figure 2(d). It is shown that the intensities of the peaks of (002) and (102) are enhanced with the increasing of pressure. At a hot pressing pressure of 8 MPa, the intensity of the peak of the (002) plane is the highest, indicating a preferential orientation along the (002) crystal plane. This suggests that the application of uniaxial pressure perpendicular to the film surface induces preferential orientation along the (002) crystal plane when large grains undergo deformation or fracture. From the MD simulation results in figure 5, it is evident that the (002) crystal plane exhibits a significantly high thermal conductivity (details will be discussed in the subsequent section on MD simulations). Therefore, orienting along the (002) crystal plane will enhance the thermal conductivity of the PCs film and induce anisotropy.

It is worth noting that Harada et al. found that for $[(CH_3)_4N][FeCl_4]$ films, the peak of the (002) crystal plane in Powder-XRD disappears when the temperature rises to 360 K[11]. Additionally, as shown in Figure 2(c) of this study, when the temperature increases to 372 K, the in-plane thermal conductivity and anisotropy of the film decrease significantly, further demonstrating the important relationship between crystal orientation and anisotropy.

However, when the pressure exceeds 8 MPa, the intensities of the (002) and (102) diffraction peaks markedly decrease. This indicates that excessive pressure partially disrupts the orientation along the (002) and (102) directions, leading to a decrease in thermal conductivity. However, as shown in Figure 2(a), a slight increase in thermal conductivity is observed. As shown in Figure 4(d), this is attributed to the pressure-induced formation of distinct layered structures in the film (details will be discussed in the subsequent section on cross-sectional SEM characterization).

To better elucidate the impact of hot pressing on thermal conductivity, the morphology of the PCs films was characterized by scanning electron microscopy (SEM). The surface morphology of films is shown in Figure 3. With the increase of pressure, the film surface progressively flattens, and inter-grain bonding improves. At 4 MPa pressure, the film surface exhibits protruding island-like grains. With increasing pressure, the island-like structures diminish, leading to a flatter surface of the film. Gaps between grains are compressed, resulting in closer contact between grains. Consequently, the number of grain boundaries in the in-plane direction decreases, which contributes to enhanced phonon transport in the in-plane direction. Consequently, this results in an increase in the in-plane thermal conductivity of the PCs film.

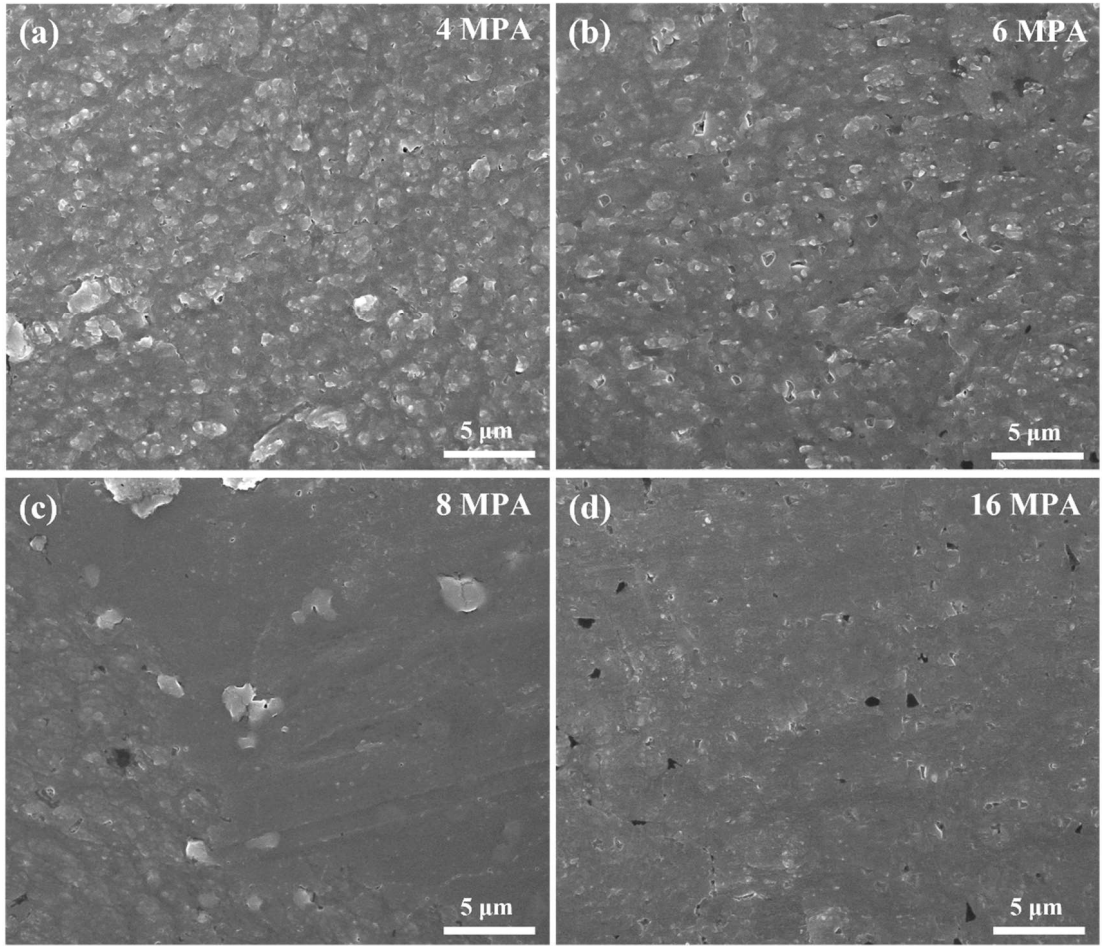

Figure 3: SEM surface morphology of PCs films prepared under varying pressures: (a) 4GPa, (b) 6 MPa, (c) 8 MPa, and (d) 16 MPa.

The cross-sectional SEM topography in Figure 4 offers visual confirmation of the selective crystal orientation and a high anisotropy in thermal conductivity. Figure 4 illustrates the progressive deformation of grains within the film as pressure increases, leading to the development of a layered structure. Throughout the hot pressing process, the expulsion of air from interstices between grains and within grain boundaries results in reduced gaps. This closer grain-to-grain contact yields a denser film, facilitating enhanced phonon transport. At a hot pressing pressure of 4 MPa, the pressure induces a tilted state in part of the grain boundary, and there are still some substantial pores present between grains. Upon increasing the pressure to 6 MPa, laminated grains begin to manifest within the film. Further elevation of pressure to 8 MPa results in robust interconnection of grains along the in-plane direction, accentuating the emergence of a more distinct layered structure. This phenomenon may arise due to the pressure conditions favoring the interconnection of grains along the in-plane direction, consequently enhancing the thermal transport capacity within the plane. This results in the crystal exhibiting preferential orientation along the (002) crystal plane. It is worth noting that when the pressure reaches 16 MPa, a distinct layered structure appears within the film. This is attributed to the gradual compression of grains into elongated shapes under increasing uniaxial pressure, leading to tight interconnections between grains along the in-plane direction and the formation of a layered structure. The emergence of this layered structure significantly elevates the thermal transport performance in the in-plane direction.

It is noteworthy that, for the cross-plane direction, with increasing pressure, the compression of grains results in a higher number of grain boundaries per unit length. Additionally, the fragmentation of grains generates additional grain boundaries. Therefore, under the influence of pressure, the film gradually forms a layered structure in the cross-plane direction. This inhibits thermal transport along the cross-plane direction, resulting in a reduction in cross-plane thermal conductivity. Therefore, under the influence of hot pressing, the crystals in the PCs film undergo preferential orientation along the in-plane direction, leading to the emergence of a layered structure. This results in a significant anisotropy in thermal conductivity of the film.

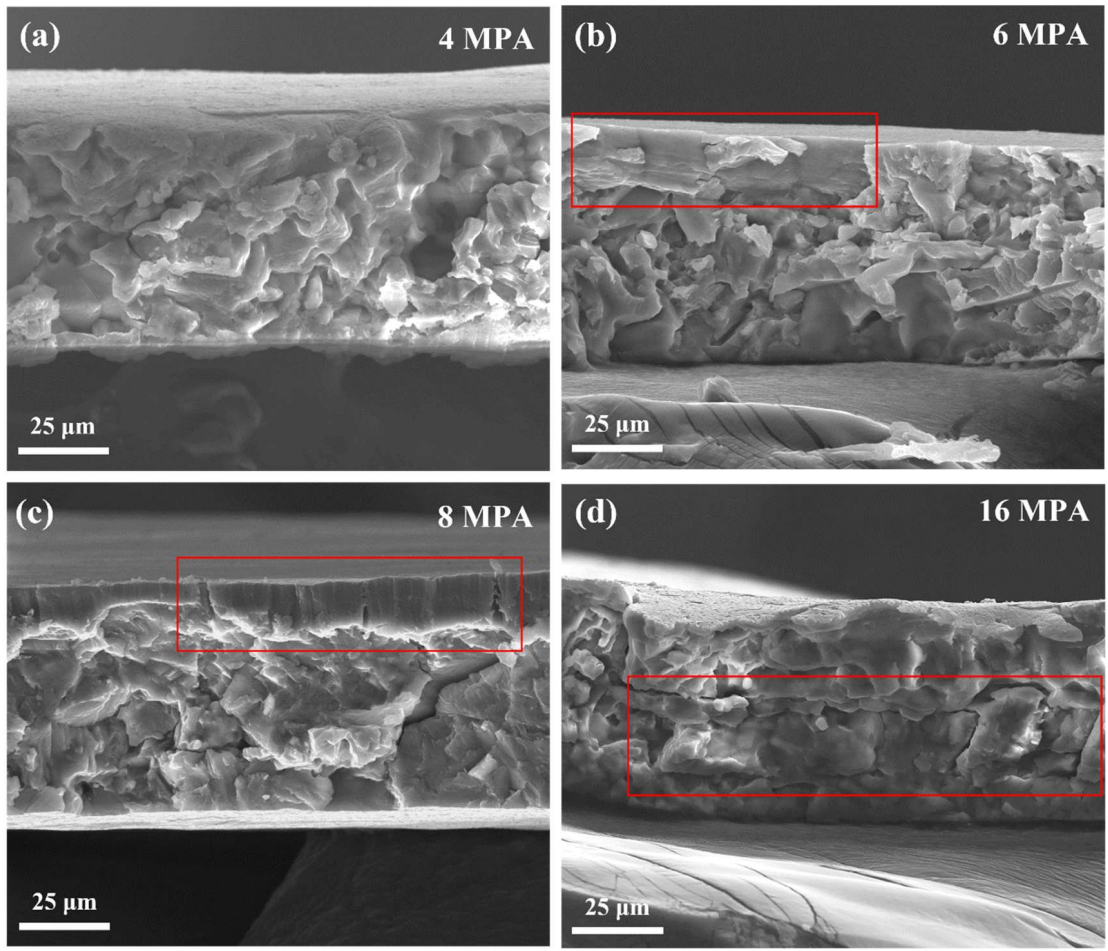

Figure 4: SEM cross-sectional morphology of PCs films under varying pressures.

However, at a pressure of 16 MPa, noticeable fractures occur within the layered region, indicating that the thermal conductivity along the in-plane direction becomes constrained when the pressure reaches a certain threshold. Consequently, as depicted in Figure 2(a), the in-plane thermal conductivity of the film pressed at 16 MPa exhibits virtually no further increase.

In addition to investigating the anisotropy in thermal conductivity, MD simulations were conducted to explore the intrinsic thermal conductivity of $[(CH_3)_4N][FeCl_4]$ along a couple of directions. The simulation results (figure 5) reveal that the intrinsic thermal conductivity of Phase III has distinct values along different directions. The highest value is $0.39 \pm 0.09$ W/m-K along [100] directions, which is close to 7.8 times higher than those along [011] directions (0.05 ± 0.02 W/m-K). Combining with Figure 2(d) (XRD images), it is evident that the crystals preferentially orient along the (002) plane. The highly conductive directions [100] and [010] are parallel to the (002) plane, indicating that the thermal conductivity along the (002) plane is stronger compared to other crystallographic directions. Additionally, Figure 2(a) shows that when the pressure increases to 8 MPa and 16 MPa, the ratio of thermal conductivity between in-plane and cross-plane directions is 5.1 and 5.5, respectively. Therefore, the (002) plane likely corresponds to the in-plane direction. In summary, at high temperatures up to 150°C, the film is in the plastic crystal phase. With the application of uniaxial pressure, the crystals preferentially orient along the high thermal conductivity (002) crystal plane, forming a layered structure in the film. This significantly enhances the in-plane thermal conductivity of the film, demonstrating an ultra-high anisotropy in thermal conductivity.

Essentially, the anisotropy in thermal conductivity comes from the different atomic couplings along different directions. In ionic PC, Coulombic forces play a dominant role, and $[(CH_3)_4N][FeCl_4]$ experiences different Coulombic forces along the [100], [010], [001], [110], [101], and [011] directions. For Phase III, it was found that by computing the forces acting on all atoms in the system along the [100], [010], [001], [110], [101], and [011] directions, the Coulombic force along the [100] direction is 5.3 times stronger than that along the [101] and [011] directions. Consequently, this results in significant differences in thermal transport capabilities along different directions. Details of simulations are in SM III.

At 360 K (Phase III), the anisotropy in thermal conductivity of $[(CH_3)_4N][FeCl_4]$ diminishes. Compared to Phase II, the thermal conductivity along the [110] and [101] directions in Phase 3 increased by 83% and 129%, respectively, resulting in a reduction in its overall anisotropy. Therefore, this phenomenon effectively explains the results depicted in Figure 2(c), where with increasing temperature, the discrepancy in intrinsic thermal conductivity values among different directions diminishes, and the rotational motion of molecules becomes more pronounced, weakening their original orientation, thereby resulting in the film's thermal conductivity exhibiting little directionality.

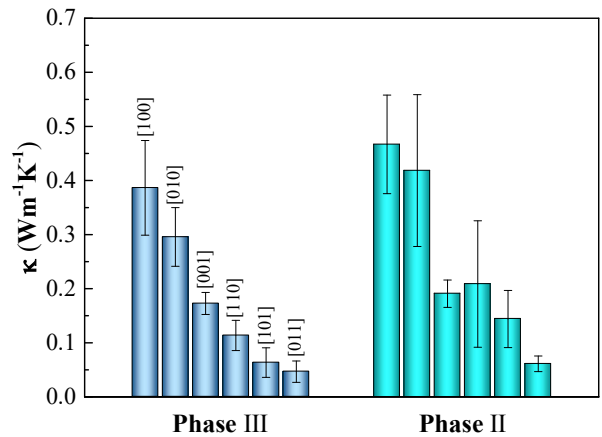

Figure 5: The values of thermal conductivity of $[(CH_3)_4N][FeCl_4]$ along different directions by MD simulations: Values for Phase III (corresponding to the temperature spread as 309K < T < 350K), and Phase II (350K < T < 385K).

MD simulations were employed to compute the thermal conductivity of single crystals under 0 MPa conditions, yielding results higher than experimental measurements on polycrystalline films, which is a reasonable outcome. However, at 8 MPa and 16 MPa, the experimental values exceeded the results of the simulations. This disparity may stem from the inability of conventional force fields to accurately describe the thermal properties of the system. Furthermore, the lack of suitable potential functions for these ionic crystal plastic crystals has made qualitative studies a focal point in this aspect of molecular simulations.

## 4. Conclusion

In summary, the thin film of [(CH$_3$)$_4$N][FeCl$_4$] plastic crystal, synthesized through uniaxial pressure hot pressing, demonstrates an ultra-high anisotropy in thermal conductivity at extremely low pressures. Specifically, under an increased pressure of 16 MPa, the in-plane thermal conductivity of the film can reach 0.59 W/m-K, with an exceptionally high ratio of in-plane to cross-plane thermal conductivity, reaching 5.5. This is attributed to the outstanding mechanical deformability of [(CH$_3$)$_4$N][FeCl$_4$] in its PC phase, which allows it to exhibit a preferred orientation along the (002) crystal plane under uniaxial pressure. This leads to the formation of a layered structure, resulting in a smooth and dense film surface. Additionally, in phase III, the thermal conductivity along the [100] and [010] directions (parallel to the (002) crystal plane) surpasses that along the [001] direction. Consequently, [(CH$_3$)$_4$N][FeCl$_4$], when hot-pressed into a film, exhibits a remarkable anisotropy in thermal conductivity at room temperature. The ability to achieve the substantial regulation of anisotropy in thermal conductivity with minimal pressure holds promise for expanding the applications of PCs films in flexible electronic devices and thermal switches.

# Conflicts of interest

**Authorship contribution statement**: Zhipeng Wu: Investigation, Writing - original draft, Data curation, Formal analysis. Mingzhi Fan: Investigation, Writing - original draft, Data curation. Yangjun Qin: Investigation, Writing - original draft, Software. Guangzu Zhang: Project administration, Conceptualization,Writing - review & editing. Nuo Yang: Project administration, Conceptualization, Writing - review & editing.